\documentclass[11pt,a4paper,amsfont]{JHEP3} 
\usepackage{epsfig}  
\preprint{\vbox{{\hbox{SOGANG-HEP 307/03}
                 \hbox{gr-qc/0307034}}}}
\title{Traversable Wormholes Construction in 2+1 Dimensions}
\author{Won Tae Kim\\Department of Physics and Basic Science Research
  Institute, Sogang University, C.P.O. Box 1142, Seoul 100-611, Korea\\
         E-mail:\email{wtkim@mail.sogang.ac.kr}}
\author{John J. Oh \\Department of Physics and Basic Science Research
  Institute, Sogang University, C.P.O. Box 1142, Seoul 100-611,
  Korea\\ E-mail:\email{john5@string.sogang.ac.kr}}
\author{Myung Seok Yoon \\Department of Physics and Basic Science Research
  Institute, Sogang University, C.P.O. Box 1142, Seoul 100-611,
  Korea\\ E-mail:\email{younms@string.sogang.ac.kr}}
\date{\today}
\abstract{
We study traversable Lorentzian wormholes in the three-dimensional low
energy string theory by adding some matter source involving a
dilaton field. It will be shown that there are two-different types of wormhole 
solutions such as BTZ and black string wormholes 
depending on the dilaton backgrounds, respectively. We finally obtain the
desirable solutions which confine exotic matter near the throat of
wormhole by adjusting NS charge.}
\begin{document}
\maketitle
\section{Introduction}
Since Morris and Thorne have verified the realistic
possibilities of constructing a traversable wormhole spacetime and
traveling through it in the theoretical context of 
the general relativity \cite{mt}, many scientists have had great interests
in this curious and weird object even though they still wonder if it
can be used for a miraculous but wishful transportation of
interstellar journey in our world or backward travel to past world.

Topologically, wormhole spacetimes are the same as that of black
holes, but a minimal surface called {\it throat of wormhole} is
maintained in time evolution, and then a traveler can pass through it
in both directions. To support a throat of wormhole, an extra matter
source called {\it exotic matter} should be added to the Einsten's
equation, which inevitably violates energy
conditions including weak(WEC), strong(SEC), and dominant energy
conditions (DEC). It is believed that the fact that the energy density
of every kinds of matter should be non-negative everywhere is restricted to the
classical system. As for {\it exotic matter}, it might let us give up
all powerful theorems such as the singularity theorem and the positive
mass theorem that require some types
of energy conditions in the classical theory of gravity. Visser et
al. showed that traversable wormholes necessarily require a violation
of the averaged null energy condition (ANEC) and can be supported by
arbitrarily small amount of {\it exotic matter}~\cite{vis1}. And many
authors intensively have studied on various aspects of 
the energy condition of traversable
wormholes in Refs.~\cite{vis2,bv,ih,and}. 

On the other hand, the string theory unlike the general relativity is
described by the gravity coupled to dilaton field and additional
gauge fields carrying Neuve Schwarz(NS), Ramond(R) charges, and so
on. Especially in 2+1 dimensions, the geometric solutions 
are well-known Ba\~nados-Teitelboim-Zanelli(BTZ) 
black hole \cite{btz} for the constant dilaton field, and the other is the
black string solution for the logarithmic dilaton which is 
dual to the constant dilaton \cite{hw}. Now, it will be interesting to study
a traversable wormhole in the the string theory, 
following the conventional recipe by introducing
an additional {\it exotic} matter source to support a throat of
wormhole. However, in string theory compared to the
conventional Einstein theory, the consistency of the equation of motion
makes the additional matter source extend non-trivially along with 
the dilaton field, which will be shown in later.
So we shall assume that the total action can be written in the form of 
the string theory with an additional
matter action as
\begin{equation}
  \label{actiond}
  S_{\rm total} = S_{\rm string}(g,\phi, B_{\mu \nu}) + S_{\rm M}(f,g,\phi),
\end{equation}
where $g$ is a metric, $\phi$ is a dilaton field, $B_{\mu \nu}$ is
an antisymmetric tensor field, and $f$ is a matter field. The additional
matter part $S_{\rm M}$ contributes to each equation of motion for
$g_{\mu\nu}$ and $\phi$ to form a wormhole, which determines
the desirable wormhole solutions depending on some matter distributions. 
Note that the traversable wormholes in three-dimensional Einstein
gravity with a cosmological constant are intensively studied in
Ref. \cite{dm}.

In this paper, some general conditions of traversable
wormholes are presented in Sec.~\ref{sec:II}. 
In Sec.~\ref{sec:III}, we shall obtain traversable wormhole solutions 
in the three-dimensional low energy string action 
by adding the additional matter term involving dilaton field. 
The various solutions of traversable
wormholes for two-different types of dilaton backgrounds are presented
and the {\it exotic} behaviors of 
the corresponding matter distributions are shown in
Sec.~\ref{sec:IV}. Finally, in Sec.~\ref{sec:V}  some 
discussions and comments on our results are given.  
  
\section{Preliminary: Traversable Wormhole \hfill{}\label{sec:II}} 
In this section, we shall briefly discuss some conditions for
forming a traversable wormhole and expressions of energy-momentum
tensors in the proper reference frame. 
A traversable wormhole can be constructed by assuming a metric ansatz of
spherically symmetric form as  
\begin{equation}
  \label{metric} 
  (ds)^2 = -e^{2\Phi(r)}d^2t + \frac{d^2r}{1-b(r)/r} + r^2 d^2\varphi,
\end{equation}
where $\Phi(r)$ is a redshift function and $b(r)$ is a shape function
of a traversable wormhole. One of the wormhole conditions~\cite{mt}
restricts the shape function $b(r)$ in the metric (\ref{metric}) to
satisfy a flaring-out condition as 
\begin{equation}
 \label{flaringout}
 [rb'(r)-b(r)]|_{r_{\rm throat}} < 0,
\end{equation}
where a prime denotes a differentiation with respect to $r$, and
a throat of wormhole is defined as 
\begin{equation}
  [1-b(r)/r]|_{r_{\rm throat}} = 0. \label{throat}
\end{equation}

Note that the restriction of the redshift function contains realistic
concepts of the traversable wormhole conditions for constructing
wormholes and traveling through them. Since there exists no horizon 
in the wormhole geometry, $\Phi(r)$ should be finite everywhere, and
$|\Phi(r)|\ll1$ and $|\Phi'(r)|\lesssim$ [ earth gravity ] for a
comfortable journey through the wormhole. 

In our model, the vacuum solutions are well-known black hole solutions
called BTZ and black string solutions. The relevant matter
action denoted by $S_{\rm M}$ should be added in order for creating 
a traversable wormhole. 
To simplify subsequent calculations and physical interpretations, 
we switch the coordinates to set of new orthonormal basis called {\it proper reference
frame}, which describes observers who always remain at rest in the
coordinate system satisfying
\begin{equation}
 \label{metric_tetrad}
  (ds)^2 = g_{\mu\nu}dx^{\mu}dx^{\nu} 
  = \eta_{\hat \alpha \hat \beta}dx^{\hat \alpha}dx^{\hat \beta}
\end{equation}
with
\begin{equation}
  \label{tetrad}
 {\bf e}_{\hat t} = e^{-\Phi(r)}{\bf e}_{t}, 
 \quad {\bf e}_{\hat r} = \left(1-\frac{b(r)}{r}\right)^{1/2}{\bf e}_{r}, 
 \quad {\bf e}_{\hat \varphi} = r^{-1}{\bf e}_{\varphi},    
\end{equation}
where ${\rm diag}(\eta_{\hat \alpha\hat\beta})=(-1,1,1)$. 
In this frame, we define the energy-momentum
tensor ${T}_{\mu\nu}^{\rm M}$ as 
\begin{equation}
  \label{energytensor}
  {T}_{\hat t\hat t}^{\rm M}=\rho(r), ~~{T}_{\hat r \hat
  r}^{\rm M}=-\tau(r), ~~{T}_{\hat \varphi \hat \varphi}^{\rm
  M}=p(r),
\end{equation}
where $\rho$ is the density of mass-energy, $\tau$ is the tension for
the radial direction, and $p$ is the pressure for the angular direction.
Note that these energy-momentum tensors are exactly determined by the
metric solutions satisfying the appropriate wormhole conditions as
discussed above.

\section{Construction of Stringy Traversable Wormhole \hfill{}\label{sec:III}}
Although the metric is coupled to dilaton field and gauge field in the string
theory, it seems to be difficult to construct traversable wormhole
without considering an additional ({\it exotic}) matter, which means
that those fields do not play any role in constructing the traversable 
wormhole. So we will consider additional matter by adding matter
action to the original string action, then the total action is given as
\begin{equation}
  \label{action} 
  S = \frac{1}{2\pi\ell}\int d^3x \sqrt{-g}e^{-2\phi} 
      \left[R+4(\nabla\phi)^2+\frac{2}{\ell^2}-\frac{1}{12}H^2\right] 
      + S_{\rm M}(g_{\mu\nu},f,\phi),
\end{equation}
where $\phi$ is a dilaton and $H$ is a NS field strength
with $H=dB$. Note that the stringy action in Eq.~(\ref{action}) has two
dual invariant solutions : one is the BTZ black hole \cite{btz} for
the constant dilaton, and the other is the black string solution for
the dual solution of the constant dilaton (the logarithmic dilaton)
when there exists no additional matter fields, $S_{\rm M}=0$.

Varying (\ref{action}) with respect to metric, dilaton,
and antisymmetric tensor fields yields the equations of motion 
\begin{eqnarray}
  & & G_{\mu\nu} + 2\nabla_{\mu}\nabla_{\nu}\phi -
  \frac{1}{4}H_{\mu\alpha\beta}{H_{\nu}}^{\alpha\beta} 
     - \frac{1}{2}g_{\mu\nu} \left[4\Box\phi-4(\nabla\phi)^2 
      + \frac{4}{\ell^2}-\frac{1}{12}H^2\right] 
      =T_{\mu\nu}^{\rm M}, \label{eqnmot-g} \\
  & & R-4(\nabla\phi)^2+4\Box\phi+\frac{4}{\ell^2}-\frac{1}{12}H^2 
      = {\cal F}^{\rm M}, \label{eqnmot-dil}\\
  & & \nabla_{\mu}(e^{-2\phi}H^{\mu\nu\rho}) = 0, \label{eqnmot-H}
\end{eqnarray}
respectively, where $T_{\mu\nu}^{\rm
M}\sim {\delta S_{\rm M}}/{\delta g^{\mu\nu}}$ and ${{\cal F}^{\rm
  M}}\sim {\delta S_{\rm M}}/{\delta \phi}$ are energy-momentum
tensor and {\it dilaton scalar source} from 
$S_{\rm M}$. The NS-field equation (\ref{eqnmot-H}) is
exactly solved as $H^{\mu\nu\rho}={\cal
  Q}e^{2\phi}\epsilon^{\mu\nu\rho}$ where ${\cal Q}$ is a 
NS charge. 
Therefore, the equations of motion (\ref{eqnmot-g}) and
(\ref{eqnmot-dil}) are rewritten in the new coordinate system
(\ref{tetrad}) as
\begin{eqnarray}
  & & G_{\hat t \hat t} + 2\nabla_{\hat t} \nabla_{\hat t}\phi 
      - \frac{1}{2}g_{\hat t \hat t}
      \left( 4\Box\phi-4(\nabla\phi)^2+\frac{4}{\ell^2} 
      - \frac{1}{2}{\cal Q}^2e^{4\phi} \right)=\rho, \label{eqnmot2-11} \\
  & & G_{\hat r \hat r} + 2\nabla_{\hat r}\nabla_{\hat r}\phi 
      - \frac{1}{2} g_{\hat r \hat r}
      \left(4\Box\phi-4(\nabla\phi)^2 + \frac{4}{\ell^2} 
      - \frac{1}{2}{\cal Q}^2e^{4\phi} \right)=-\tau,  \label{eqnmot2-22}\\
  & & G_{\hat\varphi \hat\varphi} 
      + 2 \nabla_{\hat \varphi}\nabla_{\hat \varphi}\phi -
      \frac{1}{2}g_{\hat \varphi \hat \varphi}
      \left(4\Box\phi-4(\nabla\phi)^2+\frac{4}{\ell^2} 
      - \frac{1}{2}{\cal Q}^2e^{4\phi} \right)=p,  \label{eqnmot2-33}\\
  & & R - 4(\nabla\phi)^2 + 4\Box\phi + \frac{4}{\ell^2} 
      + \frac{1}{2}{\cal Q}^2e^{4\phi}={{\cal F}}^{\rm M}, \label{eqnmot2-dil}
\end{eqnarray}
where the non-vanishing components of the Einstein tensor are calculated as
\begin{eqnarray}
  & & G_{\hat t\hat t} = \frac{b'r-b}{2r^3}, \label{G11} \\
  & & G_{\hat r\hat r} = \frac{1-b/r}{r}\Phi', \label{G22} \\
  & & G_{\hat \varphi \hat \varphi} = 
      \left(1 - \frac{b}{r}\right)\left[\Phi'' +
      {\Phi'}^2 + \frac{b-rb'}{2r(r-b)}\Phi'\right]. \label{G33}
\end{eqnarray}
Since Eq.~(\ref{G11}) is independent of the redshift function
$\Phi(r)$, it will determine a shape function $b(r)$ easily. However,
the other components in the Einstein tensors are associated with the
redshift function $\Phi(r)$ that should be consistently solved to
satisfy some appropriate physical conditions as discussed above. In three dimensions, 
since the ($\hat r\hat r$)-component (\ref{G22}) only 
depends upon the differentiation of the redshift
function $\Phi(r)$, the wormhole conditions that $\Phi(r)$ should be
regular everywhere and sufficiently small for a comfortable journey
determine the vanishing or constant redshift solution, while this is 
somewhat different from that in $d \ge 4$
dimensions. Mathematically, partial integration of
Eqs.~(\ref{eqnmot2-22}) and (\ref{G22})
tells us that the integration of $\Phi'(r)$ 
diverges at the throat of wormholes since it includes a term of
$\ln(r-b)$, and it violates a
finiteness of the redshift function $\Phi(r)$. Furthermore, since we
expect $|\Phi'(r)|\ll1$ for no tidal forces to human body to travel
comfortably, the regular solution of $\Phi(r)$ everywhere is only
possible when $\Phi(r)=\Phi_{0}={\rm constant}$.

On the other hand, if we consider the matter action $S_{\rm M}(f,g)$
independent of the dilaton field instead of choosing $S_{\rm
  M}(f,g,\phi)$ in
Eq. (\ref{action}), the theory has no wormhole solution due to the
inconsistency of the equations of motion as far as we are interested
in the BTZ wormhole. More precisely, 
the equations of motion is given by
using the solution for the NS-field,
\begin{eqnarray}
  \label{eq:rewritteneqn}
   & & R_{\mu\nu} + 2\nabla_{\mu}\nabla_{\nu}\phi + \frac{1}{2} {\cal
   Q}^2 e^{4\phi}g_{\mu\nu} = T^{M}_{\mu\nu}, \label{gredef}\\
   & & R - 4(\nabla\phi)^2 + 4\Box\phi + \frac{4}{\ell^2} +
   \frac{1}{2}{\cal Q}^2 e^{4\phi} = 0. \label{dilredef}
\end{eqnarray}
At first sight, the exotic matter affects the geometry in
Eq. (\ref{eq:rewritteneqn}), however, as seen from
Eq. (\ref{dilredef}), 
the curvature scalar is constant
for the constant dilaton background even in the presence of the
{\it exotic} matter, whereas it is not generally constant for the assumed 
wormhole metric, Eq. (\ref{metric}). If the (non-rotating) BTZ solution is connected
with the wormhole solution in the appropriate finite region,
then one should consider the vanishing dilaton background.  
Therefore, the consistent formulation for traversable wormholes
requires some modification on the additional matter action.
 
\section{Specific Traversable Wormhole Solutions \hfil{}\label{sec:IV}}

To study a traversable wormhole alters the direction of solving
field equations from the usual way of gravitational system in that the
matter fields are determined by an appropriate wormhole metric ansatz. Once we
construct functions of metric $b(r)$ and $\Phi(r)$ so as to shape the
wormhole by the physical speculations, then we naturally obtain the corresponding
{\it exotic} or {\it normal} matter distributions. Here, we shall
exhibit some specific wormhole solutions of some physical interests. 

\subsection{Constant dilaton background (BTZ Wormhole)\hfil{}\label{subsec1}}

A simple solution in the constant dilaton field
$\phi=0$ with vanishing redshift function following the argument in
Sec.~\ref{sec:II}, 
\begin{equation}
  \label{sol1}
  b(r)=b_{0},~~\Phi(r)=0,
\end{equation}
determines the distributions of the additional matter source and the {\it
  dilaton scalar source} as 
\begin{eqnarray}
  \label{exotic1}
  & &\rho=-\frac{b_{0}}{2r^3}-\frac{1}{4}{\cal Q}^2+\frac{2}{\ell^2},\nonumber\\
  & &\tau=-p=\frac{2}{\ell^2}-\frac{1}{4}{\cal Q}^2,\nonumber\\
  & &{{\cal F}}^{\rm M}=\frac{4}{\ell^2} - \frac{b_{0}}{r^3} +
  \frac{1}{2}{\cal Q}^2.
\end{eqnarray}
The specific dimensionless parameter $\zeta$ defined as
$\zeta=(\tau-\rho)/|\rho|$ characterizes how the
{\it exotic} or {\it normal} matters are distributed \cite{mt}, and the {\it exoticity} at
or near throat of the wormhole requires to be non-negative,
$\zeta>0$. From Eq.~(\ref{exotic1}), the {\it exoticity}
reads $|\rho|\zeta=b_{0}/{2r^3}>0$ everywhere
since $r\ge r_{\rm throat}>0$. The {\it exoticity} with respect to the
radial coordinate $r$ is shown in Fig.~\ref{zeta_phi0_1} and it
converges to zero at infinity, which is somewhat tricky to thread an
another spacetime with the wormhole in the finite region by imposing the appropriate vacuum
boundary condition.

\EPSFIGURE{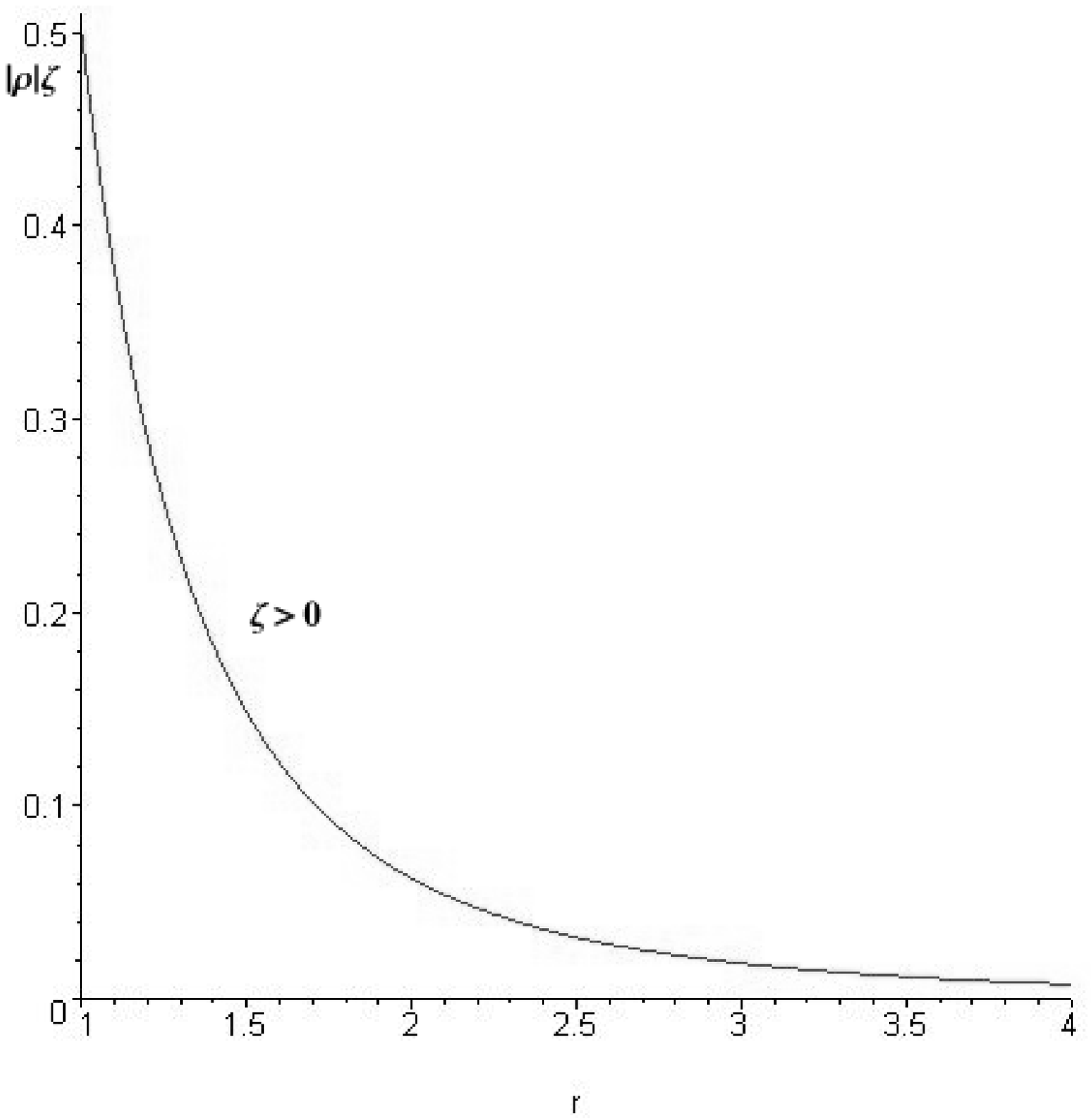, width=10cm}{\label{zeta_phi0_1}The
  {\it exoticity} $|\rho|\zeta$ for $b(r)=b_{0}$ for $b_{0}=1$. The exotic
  matter is distributed everywhere, and the asymptotic behavior shows
  that it goes to zero as $r$ increases.}
\EPSFIGURE{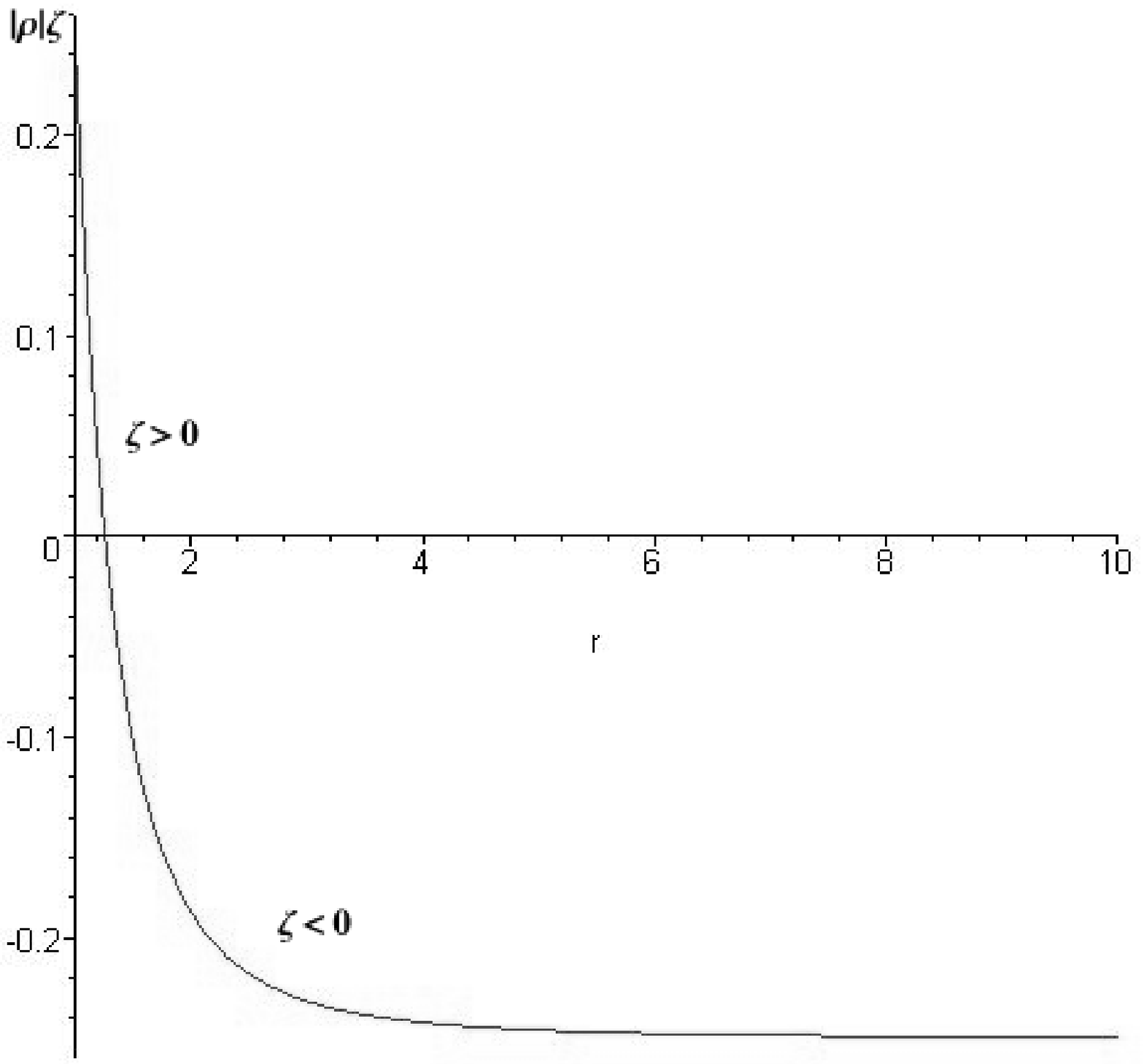,width=10cm}{\label{zeta_phi0_2}The
  {\it exoticity} $|\rho|\zeta$ for
  $b(r)=b_{0}+\frac{1}{4}{\cal Q}^2r^3$ for $b_{0}={\cal Q}=1$. The
  distribution of exotic matter is restricted to the finite region in the vicinity of
  the throat and other region is filled with the normal matters. It
  results from the NS charge contribution to the shape function $b(r)$.}

On the other hand, another class of solution of the shape function
$b(r)$ induced by the NS charge ${\cal Q}$ is obtained as 
\begin{equation}
  \label{class2sol}
  b(r)=b_{0}+\frac{1}{4}{\cal Q}^2 r^3,~~\Phi(r)=0,
\end{equation}
by choosing the additional matter fields and the {\it dilaton scalar source} as
\begin{eqnarray}
  \label{noqden}
  & &\rho=\frac{2}{\ell^2}-\frac{b_{0}}{2r^3},\nonumber\\
  & &\tau=-p=\frac{2}{\ell^2}-\frac{1}{4}{\cal Q}^2,\nonumber\\
  & &{{\cal F}}^{\rm M}=\frac{4}{\ell^2}-\frac{b_{0}}{r^3}+{\cal Q}^2.
\end{eqnarray}
The flaring-out condition (\ref{flaringout}) should be satisfied
for a traversable wormhole, and it tells us a simple
restriction between $b_{0}$ and ${\cal Q}$ at the throat of the
wormhole $r_{\rm throat}$ as $b_{0}>{\cal Q}^2r_{\rm throat}^3/2$. 
Note that it is easy to show that this flaring-out condition is equivalent to
the positiveness of the {\it exoticity}, $\rho|\zeta|_{r_{\rm
    throat}}=(\tau-\rho)_{r_{\rm throat}}>0$. Another condition of the
throat $r_{\rm throat}>0$ gives a restriction of $b_{0}$ such as
  $0<b_{0}<4/(3\sqrt{3}{\cal Q})$. Together with the flaring-out 
condition and the positiveness of the throat, we obtain a condition
between $b_{0}$ and ${\cal Q}$ as
\begin{equation}
  \label{condition1}
  \frac{1}{2}{\cal Q}^2 r_{\rm throat}^3 < b_{0} <
  \frac{4}{3\sqrt{3}{\cal Q}}.
\end{equation}
The {\it exotic} behavior of matter source can be shown by investigating
the {\it exoticity} $|\rho|\zeta=\tau-\rho=(b_{0}-{\cal
  Q}^2r^3/2)/2r^3$. Obviously, there exists a specific region where
the exoticity $\zeta$ is positive in the wormhole
spacetimes, and it means that the {\it exotic} matter can be confined
in the finite region near the throat by adding the NS charge in the
shape function $b(r)$. The behavior of the {\it exoticity} $\zeta$ is
shown in Fig.~\ref{zeta_phi0_2}, and $\zeta$ converges to 
a negative finite value $-{\cal Q}^2/4$ as $r$ goes to infinity unlike
the case of $b(r)=b_{0}$. 
Therefore, the latter case is more desirable in that the BTZ metric or
anti de-Sitter($AdS$) metric can be patched in the finite region.

\subsection{Logarithmic dilaton background (Black String Wormhole) \hfil{}\label{subsec2}}

The background dilaton $\phi=-\ln(r/\ell)$ is a dual solution
to the $\phi=0$ solution, and those two solutions
are connected by the specific symmetry called duality \cite{hw} given by
\begin{eqnarray}
  \label{duality}
  & & \bar{g}_{xx} = \frac{1}{g_{xx}}, \quad
      \bar{g}_{xa} = \frac{B_{xa}}{g_{xx}}, 
      \quad \bar{g}_{ab} = g_{ab} - \frac{1}{g_{xx}}(g_{xa}g_{xb}-B_{xa}B_{xb}),\nonumber
  \\
  & & \bar{B}_{xa} = \frac{g_{xa}}{g_{xx}},\quad \bar{B}_{ab} = B_{ab}
      - \frac{2}{g_{xx}}g_{x[a}B_{b]x},
      \quad \bar{\phi} = \phi - \frac{1}{2}\ln g_{xx},
\end{eqnarray}
where $a$ and $b$ span all directions except the compact
direction $x$.
Although the original theory have a duality as a fundamental symmetry, the same duality of
the additional matter action is not guaranteed. However, two different
dilaton
backgrounds show the
different asymptotic behaviors of vacuum solutions for $S_{\rm
  M}=0$. More precisely, the BTZ black hole($\phi=0$) shows an asymptotic
anti-de Sitter ($AdS$) behavior as $r$ goes infinity while the black
string($\phi=-\ln(r/\ell)$) has an asymptotic flat behavior. It is
expected to expose the different types of the {\it exotic matter}
distribution
 in the wormhole
spacetimes. Therefore, it will be interesting to study wormhole solutions
with this different dilaton solution regardless of preservation of
duality
symmetry.

Then, from the logarithmic dilaton $\phi= -\ln(r/\ell)$ and trivial redshift
solution $\Phi(r)=0$, the additional matter source and the {\it dilaton
  scalar source} are represented by solving equations of motion
(\ref{eqnmot-g}) and (\ref{eqnmot-dil}) in the form of
\begin{eqnarray}
  \rho &=& \frac{3}{2r^3} (b'r-b) - \frac{2}{r^2}
     \left(1-\frac{b}{r}\right) + \frac{2}{\ell^2} - \frac{{\cal Q}^2 \ell^4}{4r^4},
     \label{ldmatt_rho}\\
  \tau &=& -\frac{4}{r^2} \left(1-\frac{b}{r}\right) + \frac{2}{\ell^2} -
     \frac{{\cal Q}^2 \ell^4}{4r^4}, \label{ldmatt_tau} \\
  p &=& -\frac{1}{r^2}(b'r-b) - \frac{2}{\ell^2} + \frac{{\cal Q}^2
     \ell^4}{4r^4}, \label{ldmatt_p} \\
  {{\cal F}}^{\rm
  M}&=&\frac{3}{r^3}(b'r-b)+\frac{4}{\ell^2}+\frac{{\cal
     Q}^2}{2r^4}-\frac{4}{r^3}(r-b). \label{exscalar} 
\end{eqnarray}
At this stage, in order to obtain the shape of a wormhole, 
one can simply choose a
shape function as $b(r)=b_{0}$ following the
flaring-out condition (\ref{flaringout}). Then
Eqs.~(\ref{ldmatt_rho}), (\ref{ldmatt_tau}), (\ref{ldmatt_p}), and
(\ref{exscalar}) are reduced to
\begin{eqnarray}
  \rho &=& \frac{b_0}{2r^3} - \frac{2}{r^2} + \frac{2}{\ell^2} -
       \frac{{\cal Q}^2 \ell^4}{4r^4}, \label{rholdb0} \\
  \tau &=& -\frac{4}{r^2} \left(1-\frac{b_0}{r} \right) + \frac{2}{\ell^2} -
       \frac{{\cal Q}^2\ell^4}{4r^4}, \label{tauldb0} \\
  p &=& \frac{b_0}{r^3} - \frac{2}{\ell^2} + \frac{{\cal
       Q}^2\ell^4}{4r^4}, \label{pldb0}\\
 {{\cal F}}^{\rm M}&=&\frac{2}{k} - \frac{4}{r^2}+\frac{b_0}{r^3} +
       \frac{{\cal Q}^2\ell^4}{2r^4}. \label{F_e1}
\end{eqnarray}
Note that $|\rho| \zeta = \tau-\rho = (7b_0 - 4r)/2r^3$ from Eqs.~(\ref{rholdb0}) and (\ref{tauldb0}),
the {\it exoticity} $\zeta$ is negative within $b_0 \le r < 7b_0/4$. This means that the matter distribution is
{\it exotic} near the throat $b_0$ but normal far from the throat, which
is drastically different from the previous case of $\phi=0$ and $b(r)=b_{0}$.
The behavior of the {\it exoticity} in $r$ is shown in 
Fig. \ref{zeta_phir_1} especially for
$b_{0}=1$, ${\cal Q}=1$, and $\ell=1$, and note that the $\zeta$
converges to zero as $r$ goes to infinity.

\EPSFIGURE{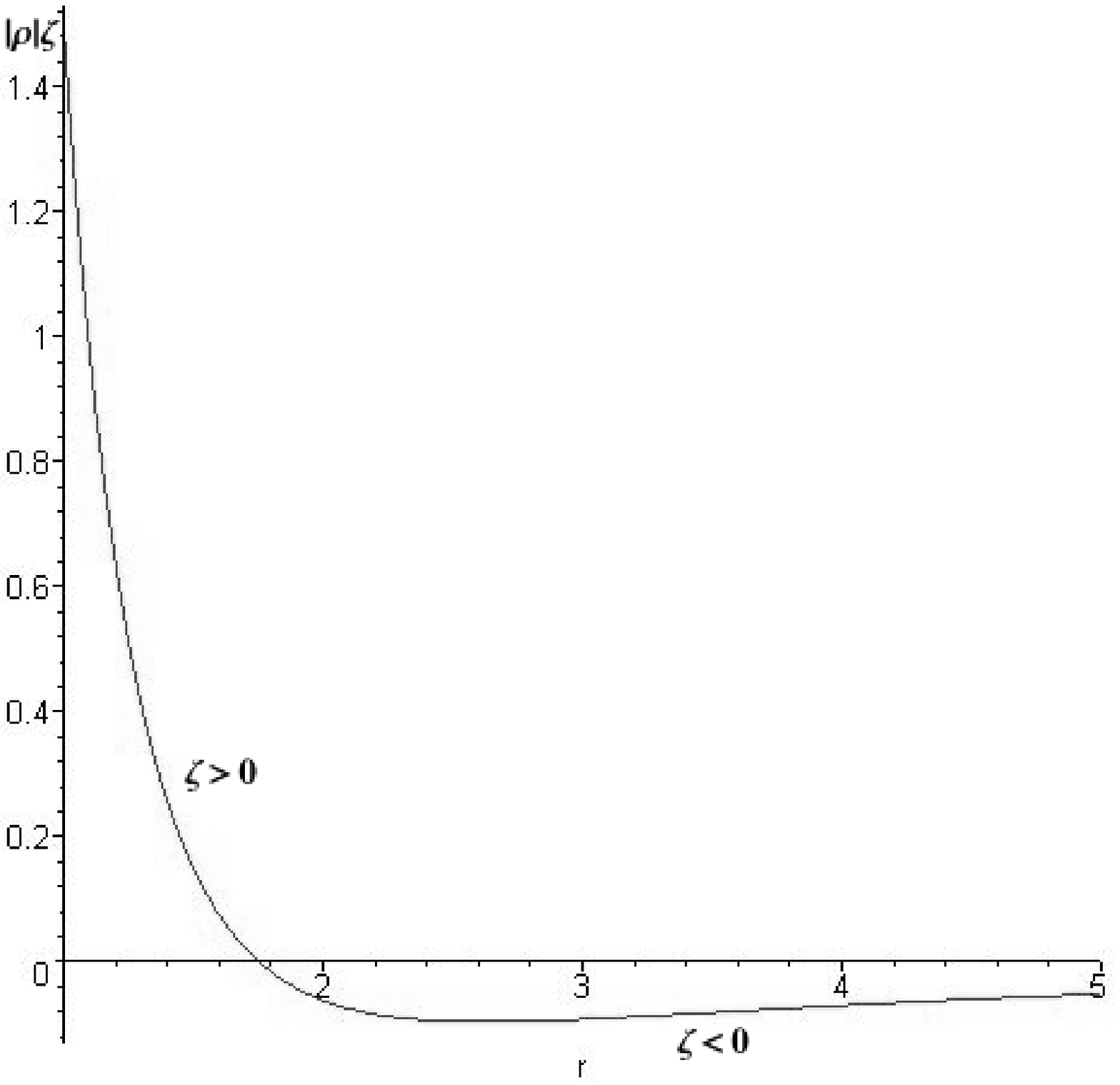,width=10cm}{\label{zeta_phir_1}The
  {\it exoticity} $|\rho|\zeta$ for
  $b(r)=b_{0}$ and $\phi=-\ln(r/\ell)$ for $b_{0}=1$ and
  ${\ell=1}$. The {\it exotic matter} is restricted to the finite
  region $b_{0}\le r < 7b_{0}/4$, which is due to the nonvanishing
  dilaton solution which is different from the constant dilaton case. }
\EPSFIGURE{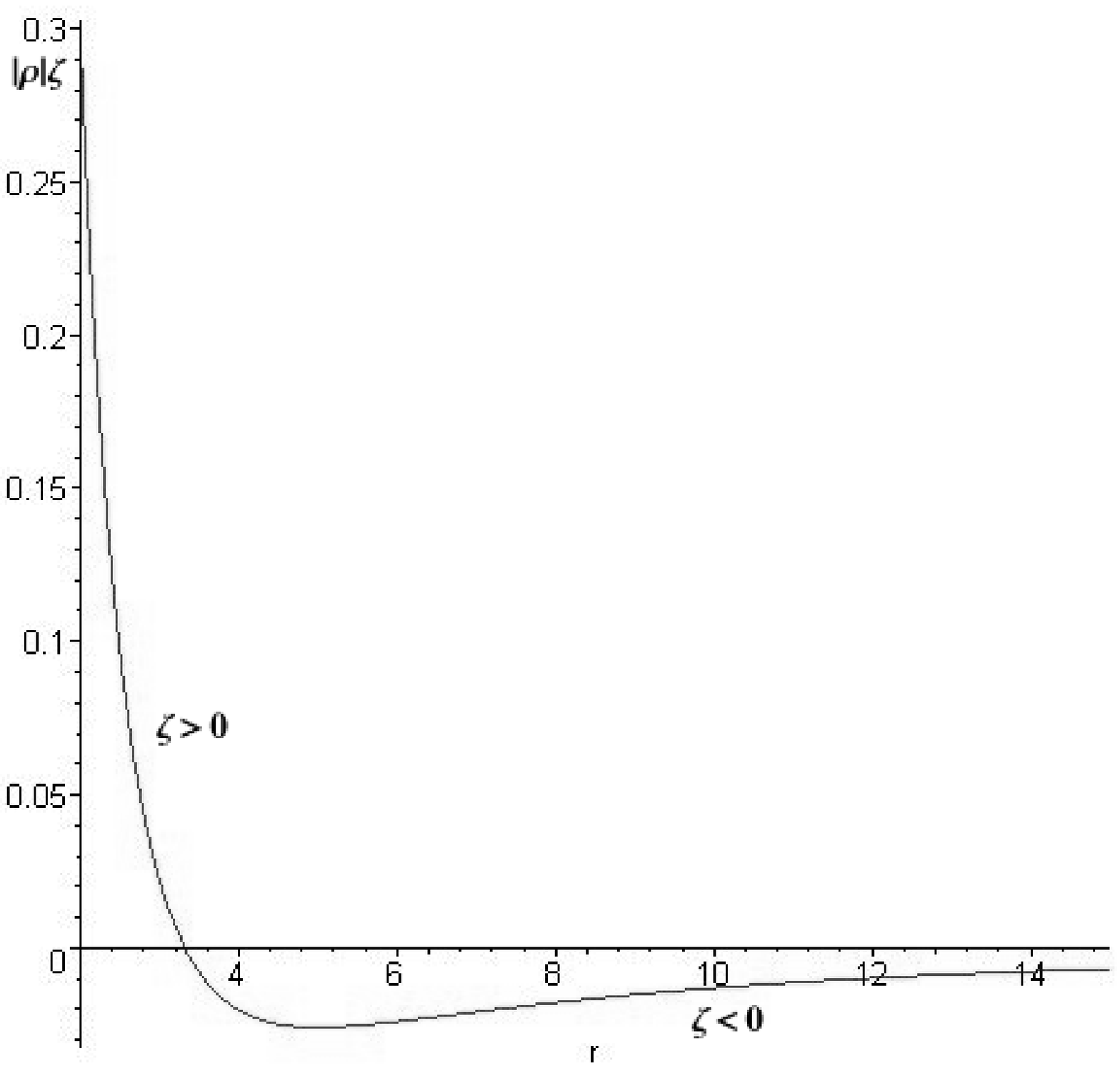,width=10cm}{\label{zeta_phir_2}The
  {\it exoticity} $|\rho|\zeta$ for
  $b(r)=b_{0}-{\cal Q}^2\ell^2/4r$ and $\phi=-\ln(r/\ell)$ for $b_{0}=3$,
  ${\cal Q}=1$, and $\ell=1$. The graph shows the similar
  behavior of $b(r)=b_{0}$ case for $\phi=-\ln(r/\ell)$.}

On the other hand, another solution including
NS charge without spoiling the flaring-out condition can be
chosen as 
\begin{equation}
  \label{eq:ldnoq}
\Phi(r)=0, \qquad b(r) = b_0 -\frac{{\cal Q}^2\ell^4}{4r},
\end{equation}
which gives the additional matter source 
and the {\it dilaton scalar source} as 
\begin{eqnarray}
  \label{eq:rhoq}
& &\rho =\frac{b_0}{2r^3} - \frac{2}{r^2} + \frac{2}{\ell^2},\nonumber
\\
 & & \tau = - \frac{4}{r^2}\left(1 - \frac{b_0}{r}\right) +
      \frac{2}{\ell^2} - \frac{Q^2\ell^4}{2r^4},\nonumber \\
  & & p = \frac{b_0}{r^3} - \frac{2}{\ell^2} - \frac{Q^2\ell^4}{4r^4},
  \nonumber \\
& &{{\cal F}}^{\rm M}=\frac{2}{\ell^2} - \frac{4}{r^2}+\frac{b_0}{r^3} + \frac{({\cal
  Q}^2\ell^4)}{r^4}.
\end{eqnarray}
The throat of the wormhole is defined at $r_{\rm
  throat} = b_0/2+\sqrt{b_0^2/4 - {\cal Q}^2\ell^4}$ for
$b_0 \ge 2{\cal Q}\ell^2$ from Eq. (\ref{eq:ldnoq}).
It is easy to show that $b(r)$
satisfies the flaring-out condition~(\ref{flaringout}) everywhere
since $r \ge r_{\rm throat}$. In this case, the {\it exotic matter} 
exists only near the throat and the {\it normal matter} is distributed
in the region of $r \ge 7b_0/8 + \sqrt{49b_0^2-40{\cal Q}^2\ell^4}/8$,
which means that the {\it exotic matter} is also confined near the throat,
and those behavior is plotted in Fig. \ref{zeta_phir_2}. 
After all, as seen from Figs. (3) and (4), the wormhole geometry with
the nonvanishing dilaton solution can be
patched onto the black string solution at the vanishing exoticity,
which is now discussed in the next section. 

\section{Discussions \hfil{}\label{sec:V}}

We have studied traversable wormholes in the three-dimensional low energy
string theory by adding the additional matter source involving the
dilaton. The key ingredient to our
traversable wormholes is that the dilaton equation should be modified by
adding the {\it dilaton scalar source}, ${{\cal F}}^{\rm M}$ for preserving
consistencies of the equations of motion.  
The appropriate physical conditions 
such as flaring-out condition for the shape function and no horizon
condition for redshift function determine the exact forms of the
additional {\it energy-momentum tensor},
${T}_{\hat\mu\hat\nu}^{\rm M}$ and the {\it dilaton scalar source}
${{\cal F}}^{\rm M}$. We require the {\it exotic matter} to be confined within a small region
around the throat, and it is surrounded by the {\it non-exotic}({\it
  normal}) matter ($\zeta \le 0$), which is considered as the best way to
minimize the use of {\it exotic matter} \cite{mt}. 
Here, we can confine our wormhole solutions to
the interior of a spherical region with a radius $r=R_{S}$, which is
corresponding to vanishing the {\it exotic} matters $\rho$, $\tau$, $p$,
and ${{\cal F}}^{\rm M}$ at all regions of $r > R_{S}$. Therefore, the
solutions with $\rho=\tau=p={{\cal F}}^{\rm M}=0$ should be naturally
patched onto outer regions at $r>R_{S}$. A simple way to thread the
whole geometry is to choose wormhole spacetimes at $r<R_{S}$, and a
(non-rotating) BTZ black hole for $\phi=0$ or a black
string for $\phi=-\ln (r/\ell)$ at $r>R_{S}$, as seen from 
Figs. (2), (3), and (4).

As for the role of the dilaton and NS charge,
for the solution of $b(r)=b_0$, the confinement of the
exotic matter near the throat appears only for the logarithmic dilaton 
case in contrast to the
$\phi=0$ case. The NS charge 
effect of the shape function $b(r)$ forms a charged wormhole, which
confines the {\it exotic matter} within the finite region near 
the throat in both dilaton cases as seen in Figs \ref{zeta_phi0_2},
\ref{zeta_phir_1}, \ref{zeta_phir_2}. Therefore, the dilaton field
$\phi$ and the NS charge contribution on the shape function $b(r)$ may play
an important role in putting the {\it exotic matter} into a finite
region in the vicinity of the throat.

On the other hand, the duality, Eq. (\ref{duality}) is considered as a
symmetry that relates a certain solution to another one in the context
of string theory. However, since it seems to be too difficult
to find an exact form of the dual invariant {\it exotic} matter action
supporting the throat of wormholes, underlying duality of this system is
manifestly broken by the specific matter distribution of forming a
traversable wormhole.

Finally, as for the two-dimensional wormholes, they apparently differ
from the higher dimensional cases in that they have a generally
covariant conformal ghost matter as an {\it exotic} matter source and
those equations of motion are exactly soluble in the conformal
coordinate system. Moreover, even though the two-dimensional model is
coupled to the dilaton field, it is unnecessary to introduce the {\it
  dilaton scalar source} because of the conformal symmetry of the
equations of motion. The wormhole solutions and their dynamics from
construction to collapse to black holes in two-dimensional dilaton
gravity are intensively studied in Ref.~\cite{hkl}. However, in $d \ge
3$ dimensions, it is difficult to find an exact form of the additional
{\it exotic} matter action.
  
\acknowledgments{J.J.O. would like to thank to H.J.Lee and E.J.Son for
helpful discussions. This work was supported by the Korea Research Foundation Grant KRF-
2002-042-C00010.}

\end{document}